\documentclass[runningheads,a4paper]{llncs}

\usepackage{amssymb}

\usepackage[final]{graphicx}
\usepackage{url}
\newcommand{\keywords}[1]{\par\addvspace\baselineskip\noindent\keywordname\enspace\ignorespaces#1}

\usepackage{amsmath,amsfonts,bbm}
\usepackage{color}
\usepackage{algorithm2e}
\usepackage{algorithmic}

\newcommand{\R}{\mathbb R}

\newcommand{\M}{\mathcal{M}}
\newcommand{\FM}{\mathcal{FM}}

\newlength\myindent
\begin{document}

\mainmatter

\title{Stochastic Development Regression on Non-Linear Manifolds}

\author{Line K\"uhnel and Stefan Sommer}
\institute{Department of Computer Science, University of Copenhagen\\
\url{kuhnel@di.ku.dk},\ \url{sommer@di.ku.dk}}

\maketitle

\begin{abstract}
We introduce a regression model for data on non-linear manifolds. The 
model describes the relation between a set of manifold valued 
observations, such as shapes of anatomical objects, and Euclidean 
explanatory variables. The approach is based on stochastic development 
of Euclidean diffusion processes to the manifold. Defining the data 
distribution as the transition distribution of the mapped stochastic 
process, parameters of the model, the non-linear analogue of design 
matrix and intercept, are found via maximum likelihood. The model is 
intrinsically related to the geometry encoded in the connection of the 
manifold. We propose an estimation procedure which applies the Laplace approximation of the likelihood function. A simulation study of the performance of the model is performed and the model is applied to a real dataset of Corpus Callosum shapes.
\keywords{Regression, Statistics on Manifolds, Non-linear Statistics, Frame Bundle, Stochastic Development}
\end{abstract}

\section{Introduction}


A main focus in computational anatomy is to study the shape of anatomical objects. Performing statistical analysis of anatomical objects is however challenging due to the non-linear nature of shape spaces. The established statistical theory for Euclidean data does not directly allow us to answer questions like: How does a treatment affect the deformation of an organ? or: Is it possible to categorize sick and healthy patients based on the shape of the subject's organs?

Shape spaces are typically non-linear and often equipped with manifold structure. Examples of manifold-valued shape data include landmarks, curves, surfaces, and images with warp variation. The lack of vector space structure for manifold-valued data implies that addition and scalar multiplication are not defined. Several concepts in statistics rely on addition and scalar multiplication, these including mean value, variance, and regression models. Hence, in order to make inference on manifold-valued data, generalization of Euclidean statistical theory is necessary. 

This paper focuses on generalization of regression models to manifolds. The aim is to model the relation between Euclidean explanatory variables and a manifold-valued response. The regression model has, as an example, applications in computational anatomy~\cite{younes_evolutions_2009}. The proposed model can for example be used to analyze how age affects the shape of Corpus Callosum~\cite{geoReg}.

Several approaches have previously been proposed for defining normal distributions on manifolds~\cite{pennec_intrinsic_2006,sommer_anisotropic}. In \cite{sommer_anisotropic}, the distribution is defined based on Brownian motions in $\R^m$ and the fact that normal distributions on $\R^m$ can be defined as transition distributions of Brownian motions. The normal distribution on the manifold is then defined as the transition distribution of the stochastic development of the Euclidean Brownian motion~\cite{hsu_stocAnMan}. The proposed regression model will be defined in a similar manner. The construction can be considered intrinsic as it only depends on the connection of the manifold, e.g. the Levi-Civita connection of a Riemannian manifold. It does not rely on linearization of the manifold, and it naturally includes the effect of curvature in the mapping of the stochastic processes.


In Euclidean linear regression, the relation between explanatory variables, $\boldsymbol{X}$, and a response variable, $\boldsymbol{y}$, is modeled by an affine function of $\boldsymbol{X}$,
\begin{align}
    \boldsymbol{y} = \boldsymbol{a} + \boldsymbol{X}\boldsymbol{b} + \boldsymbol{\varepsilon}.
    \label{ereg}
\end{align}
Due to the lack of vector space structure, alternatives for modeling relations between the given variables, $\boldsymbol{X}$ and $\boldsymbol{y}$, are needed in the non-linear situation. Several ideas have previously been introduced and a selection of these will be described in Section \ref{sec:Background}.

\begin{figure}
\centering
\includegraphics[scale = 0.35]{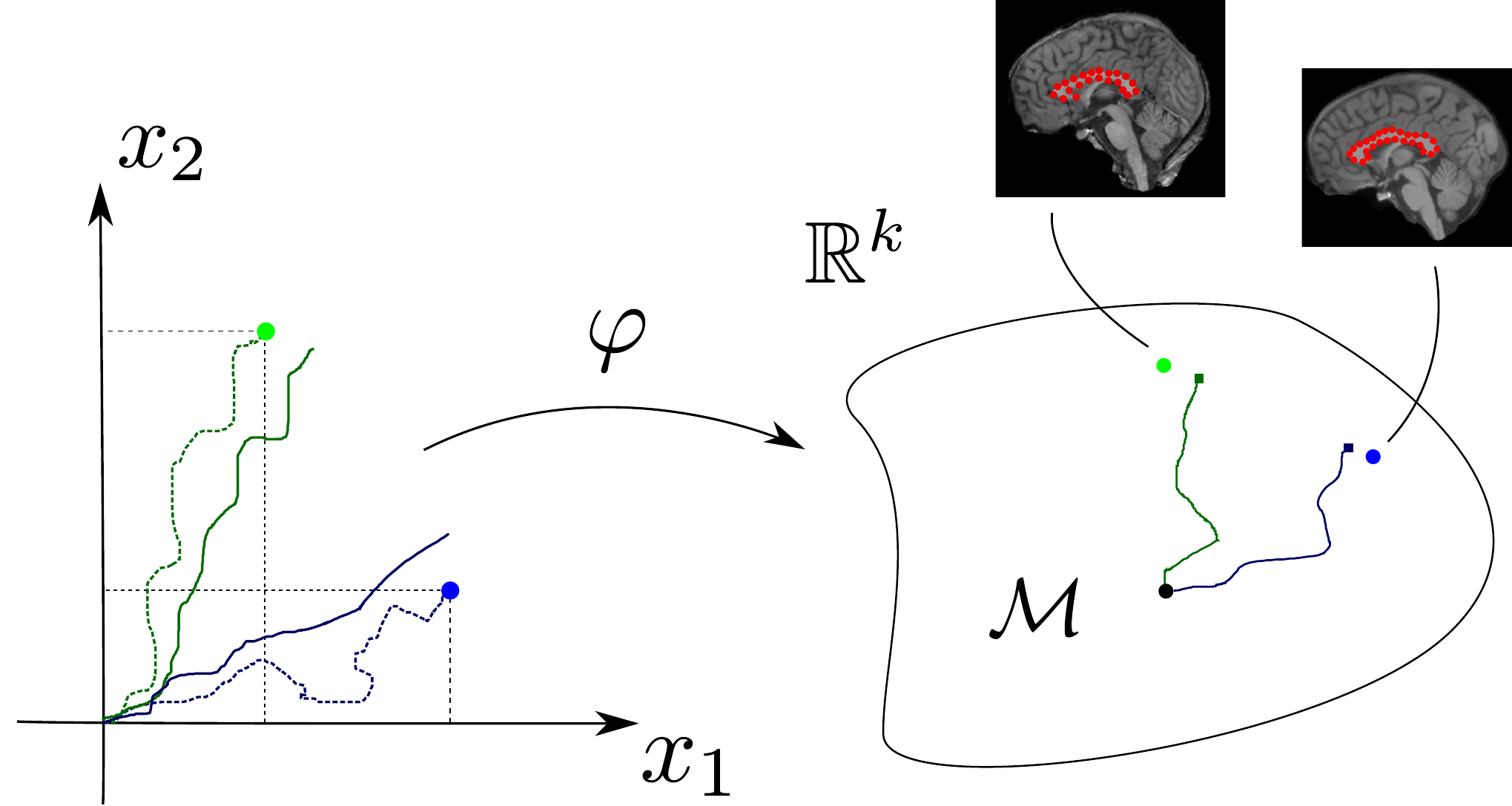}
\caption{The idea behind the proposed regression model. Stochastic processes in $\R^m$ is transported to $\M$, by stohcastic development $\varphi$, to model the relation between the explanatory variables and the response $y\in\M$.}
\label{fig:model}
\end{figure}

In this paper, the regression model is considered as a transported linear regression defined in $\R^m$. This approach is inspired by the transport of normal distributions defined in~\cite{sommer_anisotropic}. Notice that the linear regression model $(\ref{ereg})$ can be generalized to situations in which several observations are observed over time,
\begin{align}
    \boldsymbol{y}_t = \boldsymbol{a}_t + \boldsymbol{X}_t\boldsymbol{b} + \boldsymbol{\varepsilon}_t, \ \ \text{for} \ \ t\in [t_1,t_2].
\label{treg}
\end{align}
 Our approach suggests to define the regression model by transportation of stochastic processes, $Z_t = \boldsymbol{a}_t + \boldsymbol{X}_t\boldsymbol{b} + \boldsymbol{\varepsilon}_t$, in $\R^m$ on to the manifold in order to obtain the relation to the response variable, $\boldsymbol{y}$ (see Figure \ref{fig:model}).


The paper will be structured as follows. In Section \ref{sec:Background}, we give a discussion on previous methods developed for regression on manifolds. Section \ref{sec:dev} presents a short description of development of stochastic paths from a Euclidean space to the manifold. Section \ref{sec::Mod} introduces the proposed model, followed by a description of the estimation procedure in Section \ref{sec:Est}. In Section \ref{sec:sim} and \ref{sec:datEx}, illustrative examples are considered for the application and performance of the model. The paper is ended by a discussion of the defined model in Section \ref{sec:Dis}.

\section{Background}
\label{sec:Background}

Multiple approaches have been proposed for generalizing regression models to non-linear manifolds. The methods consider the regression problem in different situations. In this paper we will consider the case of Euclidean exaplanatory variables and a manifold-valued response. There have been several works describing regression models for manifold-valued data in other situations~\cite{cheng_local_2013,aswani_regression_2011,loubes_kernel-based_2009,steinke_non-parametric_2009}.


Regression models for describing the relation between a manifold-valued response and Euclidean explanatory variables have also previously been introduced. Examples include \cite{lin_extrinsic_2015} in which an extrinsic regression model is introduced, and \cite{shi_intrinsic_2009}, which defines an intrinsic regression model where the parameter vector is estimated by minimizing the total sum of squares based on the Riemannian manifold distance. Another example is the geodesic regression model introduced in \cite{geoReg}, which is a generalization of the linear regression model in Euclidean spaces. The relation is here modeled by a geodesic described by an initial velocity dependent on an explanatory variable and a starting point on the manifold.

In this paper, we will take a different view on how to relate the response and explanatory variables. Instead of considering the relation as being modeled by geodesics on the manifold as in \cite{geoReg}, we will describe the relation by stochastic paths transported from the space of explanatory variables to the manifold. By defining the regression model using stochastic paths, we are able to model non-geodesic relations, incorporate several explanatory variables, and consider random effects in the model. Non-geodesic relations have been considered by others before. An example is~\cite{singh_splines_2015} in which the geodesic regression model from~\cite{geoReg} is generalized in order to model more complex shape changes. The regression function is in this case fitted by piecewise cubic splines that describes the variation of one explanatory variable. In~\cite{hong_parametric_2016}, a regression model is introduced, in which the non-geodesic relation is obtained by time-warping. Others have proposed to model the non-geodesic relation by either a generalized polynomial regression model or by non-linear kernel-based regression~\cite{hinkle_polynomial_2012,yuan_local_2012,banerjee_nonlinear_2015,banerjee_nonlinear_2016,davis_population_2007}. On the contrary,~\cite{singh_hierarchical_2016} introduces the Hierarchical Geodesic Model which are able to consider several explanatory variables including random variables, but assumes nested observations and does only consider geodesic relations. A regression model, which incorporates both a non-geodesic relation and several explanatory variables, is proposed in~\cite{cornea_regression_2017}. This work defines an intrinsic regression model on Riemannian symmetric spaces, in which the regression function is obtained by minimizing the conditional mean of residuals defined by the log-map.


 In addition to describing the proposed model, we perform estimation of model parameters by maximum likelihood using the transition density on the manifold. The model does not linearize the manifold as in many of the local regression models, but instead take into account the curvature of the manifold at each point as encoded in the connection through the mapping of the stochastic process.

\section{Stochastic Development}
\label{sec:dev}

In this section we give a brief description of stochastic development of curves in $\R^m$ to the manifold. The reader is referred to~\cite{hsu_stocAnMan,sommer_anisotropically_2016,sommer_modelling_2016} for a deeper description of this concept.

 Let $\mathcal{M}$ be a $d$-dimensional manifold provided with a connection $\nabla$ and metric $g$. The connection is necessary for transportation of tangent vectors along curves on the manifold. A frequently used connection is the Levi-Civita connection coming from a Riemannian structure on $\M$. Let $\partial_i$ for $i=1,\ldots,d$ denote a coordinate frame on $\M$ and let $dx^i$ be the corresponding dual frame. A connection $\nabla$ is given in terms of its Christoffel symbols defined by $\nabla_{\partial_i} \partial_j = \Gamma_{ij}^k \partial_k$. For the Levi-Civita connection, the Christoffel symbols are given by
\begin{align}
    \Gamma_{ij}^k = \frac{1}{2}g^{kl}(\partial_i g_{jl} + \partial_j g_{il} - \partial_l g_{ij})
\label{eq:Chris}
\end{align}
in which $g_{ij}$ is the components of $g$ in the coordinate basis, i.e. $g = g_{ij}dx^idx^j$, and $g^{ij}$ is the inverse components.

 Consider the frame bundle $\mathcal{FM}$ being the set of tuples $(y,\nu)$ in which $y\in\M$ and $\nu$ is a frame for the tangent space $T_y\M$. Let $\pi\colon\mathcal{FM}\to\mathcal{M}$ be the projection map given by $\pi(y,\nu) = y$ for $(y,\nu)\in\FM$. A smooth curve $U_t$ on $\FM$ is a smooth selection of frames, i.e. for every $t\in I$, $U_t = (y_t,\nu_t)$ in which $\nu_t\colon\R^d\to T_{\pi(U_t)}\mathcal{M}$ is a frame.

 Given a connection $\nabla$, the tangent space of the frame bundle, $T\FM$, splits into a horizontal and a vertical part, $T\FM = H\FM\oplus V\FM$. The horizontal subspace explains infinitesimal changes of the base point on the manifold. On the other hand, tangent vectors in $V\FM$ describe changes of the frame $\nu$ keeping the base point fixed. Given a tangent vector $v\in T_y\mathcal{M}$ and a frame $\nu$, a vector in $H_{(y,\nu)}\FM$ can be defined by horizontal lift. The horizontal lift of a tangent vector $v$ is the unique horizontal vector $w\in H_{(y,\nu)}\FM$, satisfying $\pi_\star w = v$, where $\pi_\star\colon H_{(y,\nu)}\mathcal{FM}\to T_y\mathcal{M}$ is induced by the projection $\pi$. The horizontal lift of $v$ will be denoted $h_l(v)$.

Consider a probability space $(\Omega,\mathcal{F},P)$ and a stochastic process $X_t\colon\Omega\to \mathcal{W}(\R^m)$, where $\mathcal{W}(\R^m)$ denotes the path space of $\R^m$. The stochastic development of $X_t$ to $\FM$ can be defined as a solution, $U_t$, of the Stratonovich stochastic differential equation,
\begin{align}
	dU_t = \sum_{i=1}^d H_i(U_t)\circ dX^i_{t},
\label{eq:stocDev}
\end{align}
where $\circ$ symbolizes a Stratonovich stochastic differential equation. The vector fields $H_1,\ldots,H_d$ denotes a basis for the horizontal subspace of $T\FM$. Given a point $u = (y,\nu)\in\FM$, $H_i$ are defined as $H_i(u) = h_l(\nu(e_i)), \ i = 1,\ldots,d$, where $e_1,\ldots,e_d$ is the canonical basis for $\R^d$.
 A path $Y_t$ on the manifold $\M$ can then be obtained by the projection of $U_t$ onto $\M$ by the projection map $\pi$, i.e. $Y_t = \pi(U_t)$.

Consider two processes $X^1_t,X^2_t$ in $\R^m$, $t\in [0,T]$ for $T > 0$, for which $X^1_0 = X^2_0 = \boldsymbol{x}_0$ and $X^1_T = X_T^2$. If $Y^1_t$, $Y^2_t$ denotes the stochastic development of $X^1_t$ and $X^2_t$ respectively on $\mathcal{M}$, then it does not in general hold that $Y^1_T = Y^2_T$ on $\M$ due to the curvature of the manifold.

\begin{figure}
\centering
\includegraphics[scale = 0.37]{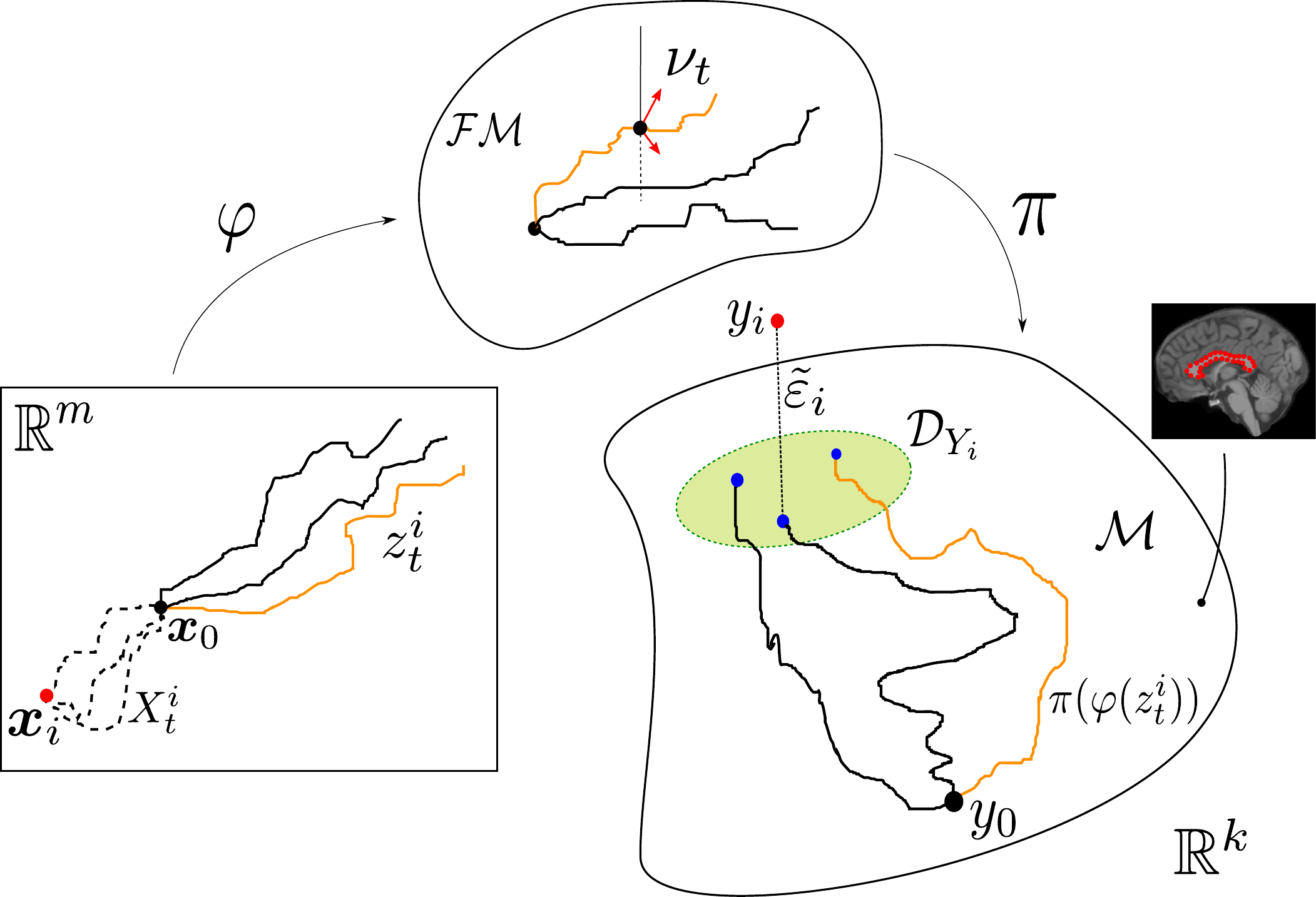}
\caption{Illustration of the regression model. Stochastic processes $z_t^{i}$, defined in $(\ref{eq:paths})$, are transported through the frame bundle $\FM$ to $\M$, with stochastic development, $\varphi$. Each observation $y_i$ is then modelled as a noisy member of the endpoint distribution of the transported $z_t^{i}$ processes. The model supports cases where the endpoint noise $\tilde{\varepsilon}$ perturbes $y_i$ in the ambient space $\R^k$ in which $\M$ is embedded.}
\label{fig:LinFig}
\end{figure}

\section{Model}
\label{sec::Mod}

Let $\M$ be a $d$-dimensional manifold embedded in the ambient space $\R^k$ for some $k\geq d$ and consider a response variable $y$ in $\M$. Let $\nu_{y_0}\colon\R^d\to T_{y_0}\M$ be a frame for the tangent space at a reference point $y_0\in\M$. Assume that $\boldsymbol{y}_1,\ldots,\boldsymbol{y}_n\in\R^{k}$ are $n$ realizations of $y\in\M$ and let $\boldsymbol{x}_i = (x^1_i,\ldots,x^m_i)\in\R^{m}$ denote the vector of explanatory variables for the $i$'th observation. Notice that the realizations of $y$ are assumed to lie in the ambient space $\R^k$ and not required to be in $\M$. This construction allows for observations measured with noise which are not necessarily observed as elements of $\M$.

The strategy of the proposed model is to define stochastic processes according to the generalized linear regression in $(\ref{treg})$ and transport these to the manifold by stochastic development. All stochastic processes are defined for $t\in [0,T]$ for a $T > 0$. Consider for each observation $i$ the stochastic process $z_t^{i}\colon\Omega\to\mathcal{W}(\R^m)$, solution to the stochastic differential equation,
\begin{align}
    dz^i_t = \beta dt + \tilde{W} dX^i_t + d\varepsilon_t.
\label{eq:paths}
\end{align}
The first term, $\beta dt$, is a fixed drift for $\beta\in\R^m$. $\tilde{W} dX_t^i$ is the dependence of the explanatory variables with $X_t^{i}\colon\Omega\to\mathcal{W}(\R^m)$ being a stochastic process satisfying $X_0^{i}(\omega) = 0$ and $X_{T}^{i}(\omega) = \boldsymbol{x}_i$ for $\omega\in\Omega$. The matrix $\tilde{W}$ is a $m\times m$-dimensional matrix with columns relating to the basis vectors of the frame $\nu_{y_0}$ on $\M$. Consider the matrix $W$ with columns consisting of basis vectors of $\nu_{y_0}$. If $\M$ has a Riemannian metric, then $W = U\tilde{W}$, in which $U$ denotes a $d\times m$ orthonormal matrix with respect to the metric.
Notice that this model can incorporate both fixed and random explanatory variables. If the $j$'th explanatory variable, $x_i^j$, is a random effect, $X_t^{ij}$ is modeled as a Brownian bridge, while it for fixed effects are modeled as a constant drift. The random error, $\varepsilon_t$, is modeled as a multidimensional Brownian motion on $\R^m$. 

The $i$'th observation $y_i$ is modeled as a noisy endpoint of the stochastic development of $z_t^i$. If $m<d$ only a reduced frame $\tilde{\nu}_{y_0}$ is used for the stochastic development of $z_t^i$. The reduced frame is considered as we are only interested in the effect of frame vectors associated to the explanatory variables. The basis vectors of $\tilde{\nu}_{y_0}$ corresponds to the columns of $W$.
Given the reference point $y_0\in\M$, define stochastic processes $Y_t^{i}$ as the stochastic development of $z_t^{i}$. Let $\mathcal{Y}^T_{i}\colon\Omega\to\M$ be a random variable following the distribution of endpoints of the stochastic development $Y_t^{i}$. Then
\begin{align}
    y_{i} = \mathcal{Y}^T_i + \tilde{\varepsilon}_{i},
\end{align}
where $\tilde{\varepsilon}_{i}\sim\mathcal{N}(\boldsymbol{0},\tau^2\mathbb{I}_d)$ 
represents the random measurement error that pulls the realization, $y_i$, from the manifold. In Figure \ref{fig:LinFig}, the two steps of the model are illustrated. First, the stochastic development of $z_t^i$ are defined on the frame bundle and finally, this stochastic development is projected to the manifold.

Notice that in the case $\M = \R^k$ with the standard connection on $\R^k$, the proposed model reduces to the regular regression model for data in $\R^k$. Assume $y\in\R^k$ and that $X_t^i$ is a vector from $0$ to $\boldsymbol{x}_i$. Then $\beta$ and $y_0$ relates to the intercept, $W$ is the matrix of regression coefficients and $\varepsilon_t$ and $\tilde{\varepsilon}$ the iid. random noise. 


\section{Estimation}
\label{sec:Est}

The reference point $y_0$, the matrix $W$, the drift $\beta$, and the variance parameter $\tau^2$ are the parameters of the model.
 These parameters can be estimated in several ways. This section describes a Laplace approximation of the marginal likelihood function which are used for finding optimal parameter estimates. We could alternatively use a Monte Carlo EM based procedure using simulations of the missing data, $Y_t^i$ for $t\in [0,T]$, to optimize the complete data likelihood. This will be considered in future works.

Laplace approximation can be used to determine a linear approximation of a non-linear likelihood function~\cite{kass_approximate_1989}. Let $\theta$ denote the vector of parameters, and $d\boldsymbol{x}_t$ a discretization of the process $X_t$ at $n_s+1$ time-points. Hence $d\boldsymbol{x}_t$ is a vector of length $n\cdot m\cdot n_s$, in which $n_s$ denotes the number of time steps, $n$ the number of observations, and $m$ the number of explanatory variables. Let $f(y\lvert\theta)$ be the conditional density of the response $y\in\M$ given $\theta$ and $p(d\boldsymbol{x}_t\lvert\theta)$ the density of the discretization of $X_t$ given $\theta$. To find the optimal parameter vector, $\theta$, the following likelihood has to be optimized,
\begin{align}
L(\theta;\boldsymbol{y}) &= f(y\lvert\theta) = \int f(y\lvert d\boldsymbol{x}_t,\theta)p(d\boldsymbol{x}_t\lvert\theta) d(d\boldsymbol{x}_t) = \int e^{-n h(d\boldsymbol{x}_t)} \ d(d\boldsymbol{x}_t),
\end{align}
where $h(d\boldsymbol{x}_t) = -\frac{1}{n} \log f(y\lvert d\boldsymbol{x}_t,\theta) - \frac{1}{n}\log p(d\boldsymbol{x}_t\lvert\theta)$.
The Laplace approximation of $L$ is then given by
\begin{align}
L(\theta;\boldsymbol{y}) \approx f(y\lvert d\boldsymbol{x}^{o}_t,\theta)p(d\boldsymbol{x}^{o}_t\lvert\theta)(2\pi)^{\frac{mn_s}{2}}\lvert\Sigma\lvert^{\frac{1}{2}} n^{-\frac{mn_s}{2}},
\label{eq:apL}
\end{align}
in which $d\boldsymbol{x}^{o}_t = \text{argmax}_{d\boldsymbol{x}_t}\{-h(d\boldsymbol{x}_t)\}$ and $\Sigma = \left(D^2 h(d\boldsymbol{x}_t)\right)^{-1}$, the inverse of the Hessian of $h(d\boldsymbol{x}_t)$. The approximated likelihood is then optimized wrt. $\theta$ to obtain the estimated parameters. In the following simulation study, the Laplace approximation is used for parameter estimation. The code for the estimation algorithm as well as the simulation study below was implemented in Theano~\cite{2016arXiv160502688short}. The code is available at \url{https://bitbucket.org/stefansommer/theanodiffgeom}.

\section{Simulation Study}
\label{sec:sim}

This section investigates properties of the model on simulated synthetic data. Two setups will be introduced, both considering landmark representations of shapes. The data are assumed to lie in a manifold defined in the LDDMM (Large Deformation Diffeomorphic Metric Mapping) framework~\cite{shapes}.

 In the LDDMM framework, deformations of shapes are modeled as smooth flows which are solutions to ordinary differential equations defined by vector fields. A point $q\in\M$ is a finite number of landmarks, $q = (x_1^1,x_1^2,\ldots,x_{n_l}^1,x_{n_l}^2)$. The metric on $\M$ is given by $g(v,w) = \sum_{i,j}^{n_l} vK^{-1}(x_i,x_j)w$, where $K^{-1}$ denotes the inverse of a kernel $K$. In this simulation study $K$ is the Gaussian kernel with standard deviation, $\sigma = 0.5$. Based on this metric the Levi-Civita connection can be obtained by calculating the Christoffel symbols defined in $(\ref{eq:Chris})$.

\begin{figure}
\centering
\includegraphics[scale = 0.35, trim = 60 30 30 30, clip]{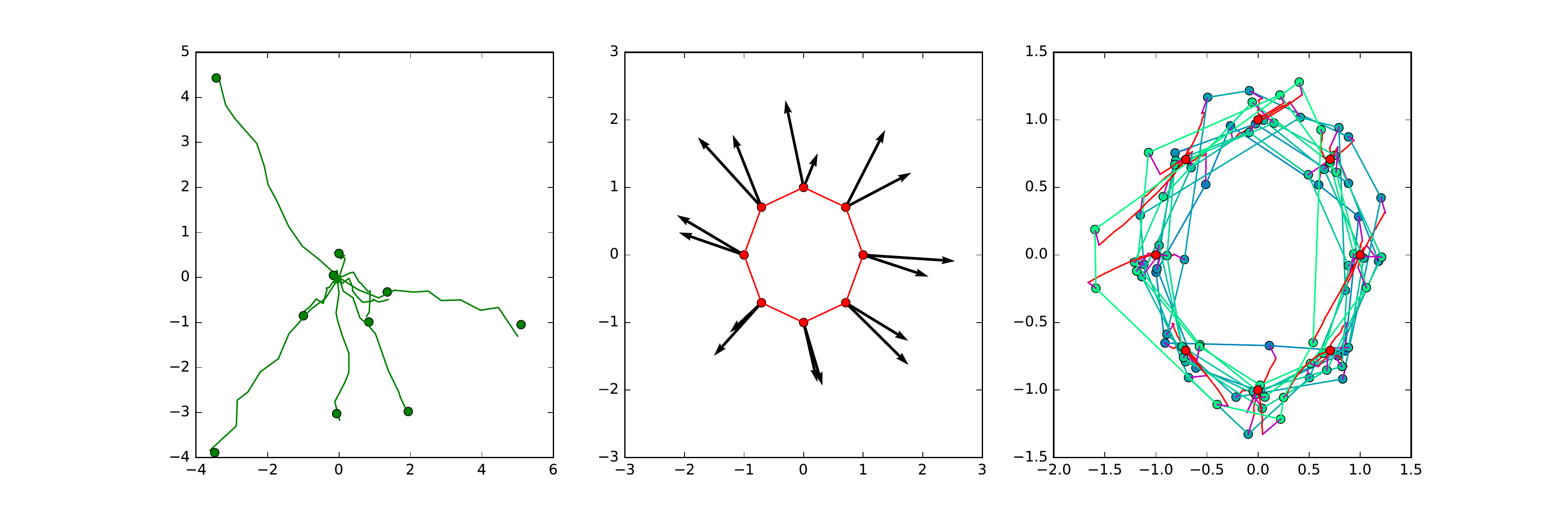}
\caption{The figures show the simulation of a dataset. (left) The stochastic paths in $\R^m$ are shown, where the vector of explanatory variables for each observation $i$ is represented by a green dot. (center) The true frame for the simulated data as well as the reference shape are plotted. (right) The simulated observations are shown, with the stochastic developments as the red processes.}
\label{fig:simCirc}
\end{figure}

To begin with, we consider estimation of $\tilde{W}$ and $y_0$ and investigate the performance of the estimation procedure. The shapes that will be considered consists of $8$ landmarks generated from the unit circle with landmarks located at $0,\frac{\pi}{4},\frac{\pi}{2},\ldots,\frac{3\pi}{2},\frac{7\pi}{4}$ radians. The center plot of Figure \ref{fig:simCirc} shows the unit circle with the chosen frame for each landmark. The number of explanatory variables are set to $m=2$ and the variables are drawn from a normal distribution with mean $0$ and standard deviation $2$. The other parameters are set to
\begin{align}
    \tilde{W} &= \begin{pmatrix}
0.2 & 0.1 \\
0.1 & 0.2
\end{pmatrix}, \ \tau = 0.1
\label{eq:simT}
\end{align}

\begin{figure}
\centering
\includegraphics[scale = 0.37,trim = 5 5 5 10,clip]{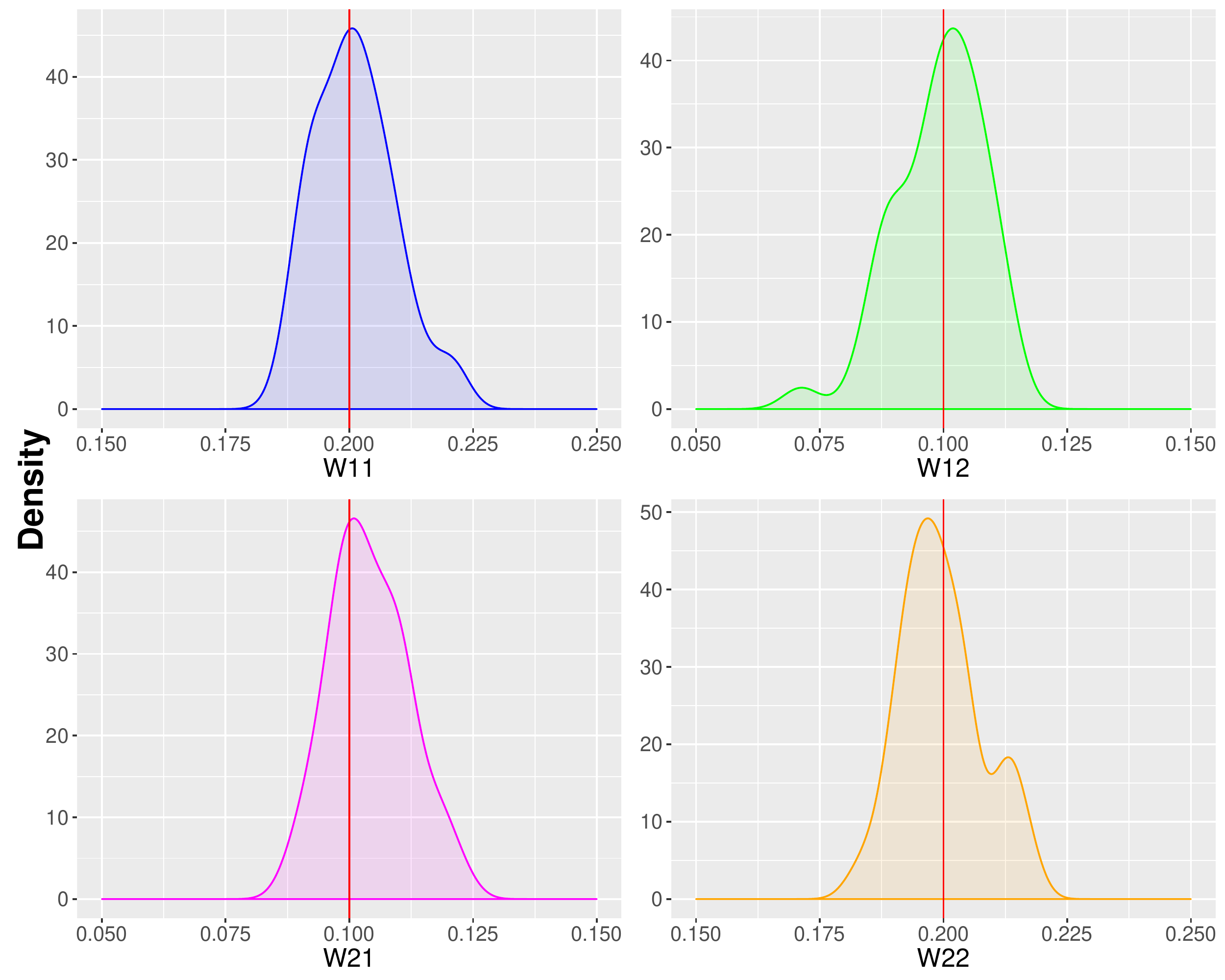}
\caption{The distribution of the estimated $\tilde{W}$ parameters. The red horizontal lines show the true parameters given in $(\ref{eq:simT})$.}
\label{fig:densFig}
\end{figure}

 In Figure \ref{fig:simCirc} is shown an example of simulated observations as well as the sample paths $X_t^i$. A total of 50 datasets were sampled, in which each consisted of 20 observations. For each simulated dataset, the $\tilde{W}$ matrix was estimated. Each of the estimated distrubtions for the entries of $\tilde{W}$ are shown in Figure \ref{fig:densFig}. By the results, we conclude that the estimated parameters are fairly stable between the different simulations and that the true values are well centered in each distribution. For this simulation, the estimation procedure is thus able to estimate the true $\tilde{W}$ parameters that were specified in the model.




\begin{figure}
\centering
\begin{minipage}{0.5\textwidth}
    \includegraphics[scale = 0.4, trim = 15 10 10 10, clip]{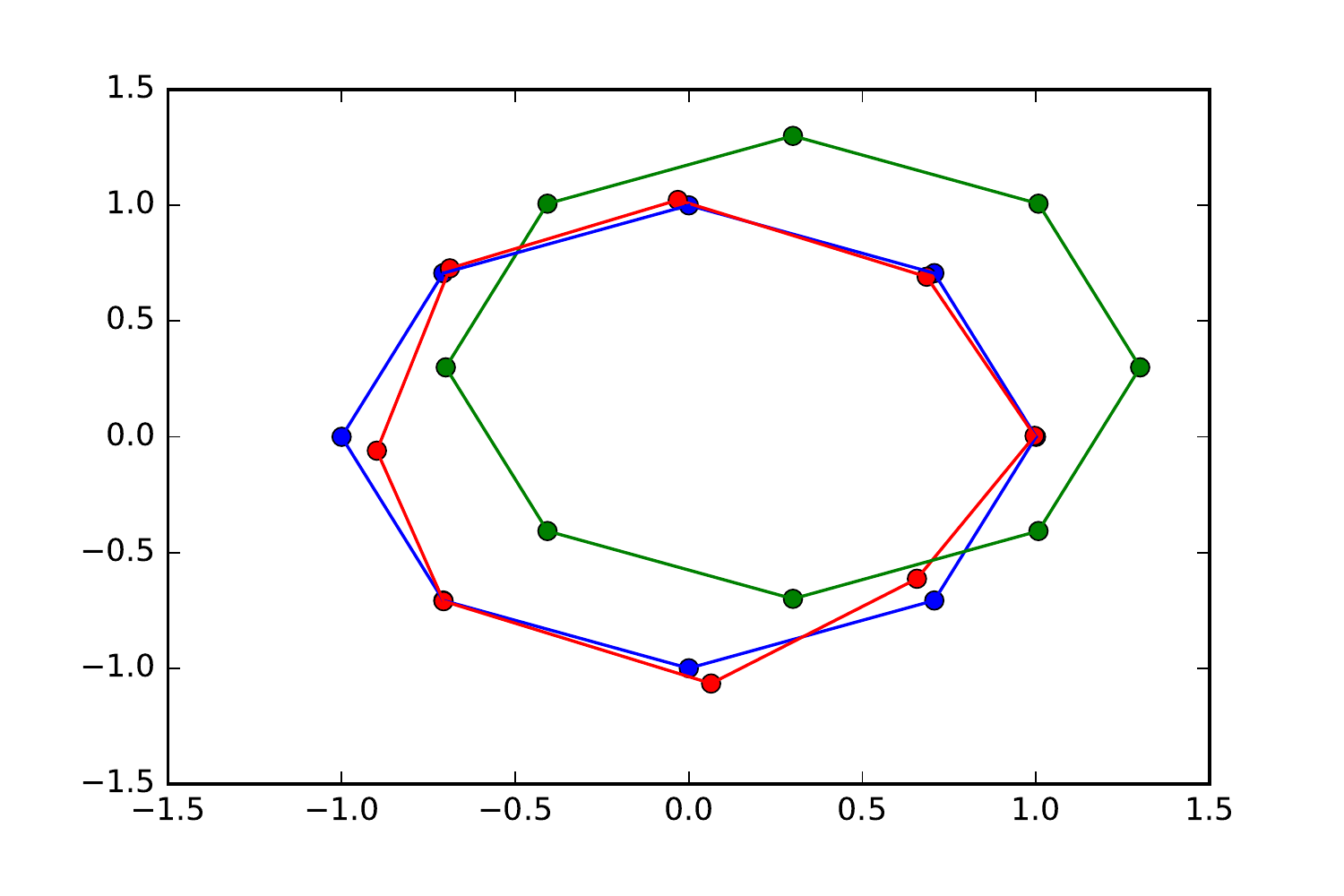}
\end{minipage}%
\begin{minipage}{0.5\textwidth}
    \includegraphics[scale = 0.4, trim = 15 10 10 10, clip]{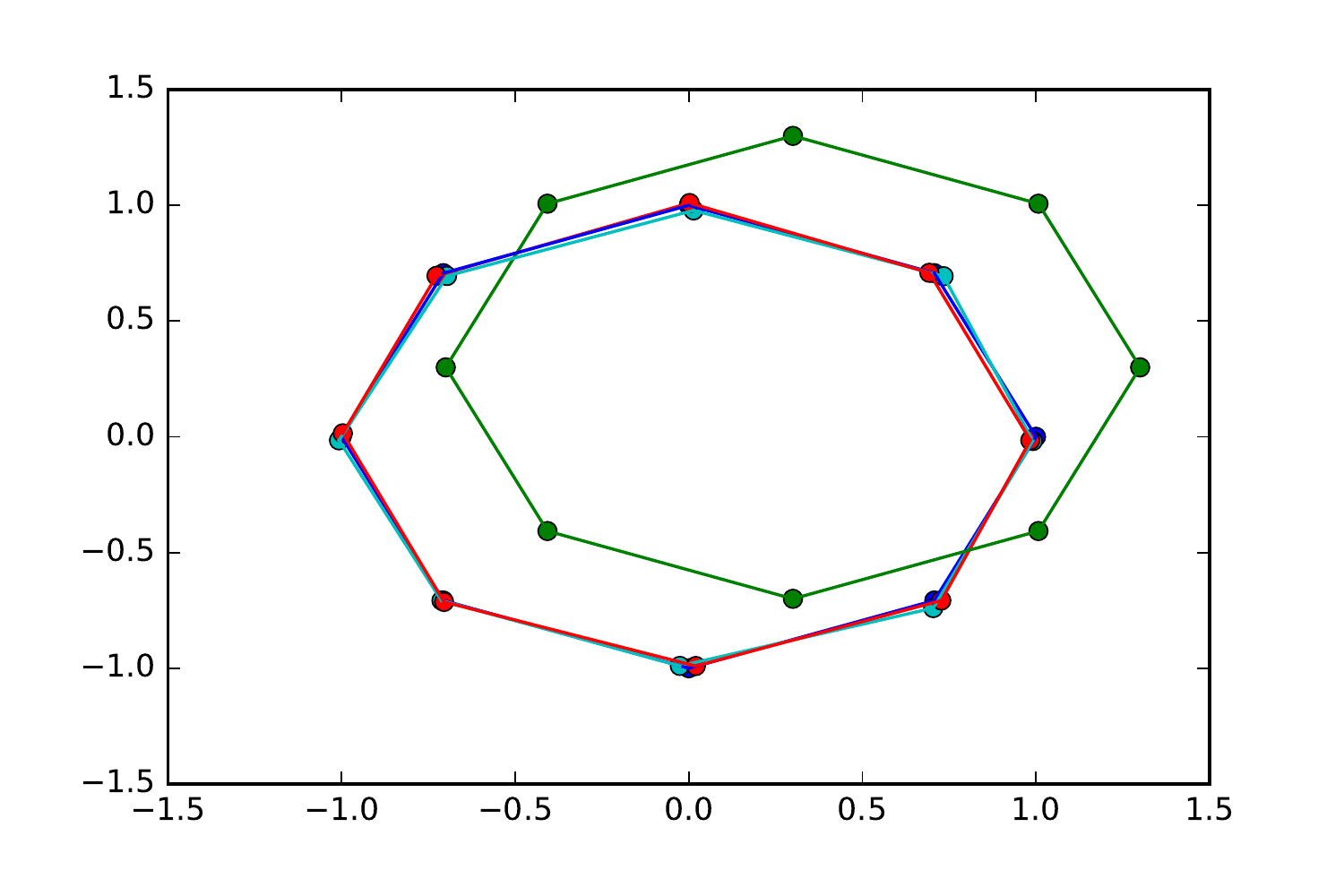}
\end{minipage}
\caption{(left) The estimated reference point $y_0$ (red) for the dataset with $20$ observations. (right) The estimated $y_0$ for $60$ (cyan) and $100$ (red) observations. In both plots, the initial (green) and the true reference circle (blue) are shown.}
\label{fig:refCirc}
\end{figure}

 Three similar datasets, as explained above, were sampled with different number of observations, 20, 60 and 100 respectively. The matrix $\tilde{W}$ as well as the reference point $y_0$ were estimated for each of the three datasets. In this case, the estimated $\tilde{W}$ matrix was found to be
\begin{align}
\hat{W}_{20} = \begin{pmatrix}
0.206 & 0.136 \\
0.147 & 0.322
\end{pmatrix}, \quad \hat{W}_{60} = \begin{pmatrix}
0.22 & 0.11 \\
0.11 & 0.21
\end{pmatrix}, \quad \hat{W}_{100} = \begin{pmatrix}
0.205 & 0.104 \\
0.115 & 0.214
\end{pmatrix}
\end{align}
while the estimated reference points are shown in Figure \ref{fig:refCirc}. By increasing the number of observations, we conclude that the estimated parameters $\tilde{W}$ and $y_0$ converge towards the true parameters.

\begin{figure}
\centering
\includegraphics[scale = 0.5, trim = 20 20 20 20,clip]{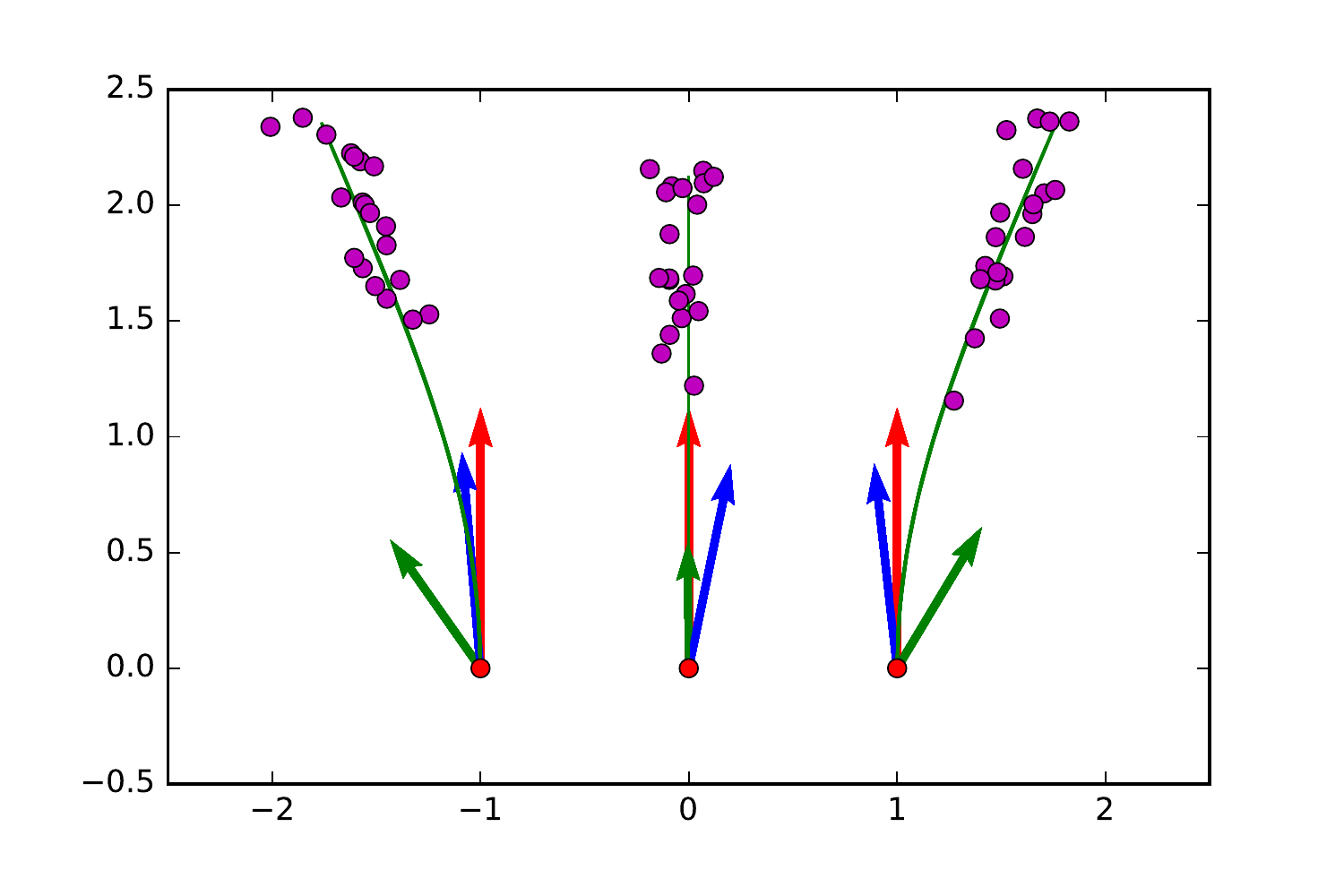}
\caption{Comparison of the estimated (blue), initial (green) and true frame vectors (red).}
\label{fig:Vec}
\end{figure}

In the second study, we consider the problem of estimating the frame matrix $U$. In this case, each observation consists of 3 landmarks that were generated from a setup shown in Figure \ref{fig:Vec}. We only consider one explanatory variable, meaning that only one frame vector has to be estimated for each landmark. The true frame vectors for each landmark was set to a vertical unit vector. In the estimation procedure, the frame vectors were initialized with the Euclidean linear regression estimate. In Figure \ref{fig:Vec} is shown the true (red), the initial (green) and the estimated frame (blue) for each landmark. The estimation procedure converges to a good estimate of the true frame. Estimation of the initial frame was considered for different number of observations, but the estimated frame did not seem to converge for increasing number of observations. The difference in the parameter estimates might therefore be a result of either the linear approximation of the likelihood or that the optimal solution of the initial frame is not unique.

\section{Data Example}
\label{sec:datEx}

We now apply the model to a real dataset consisting of landmark representations of Corpus Callosum (CC) shapes. The model is used to describe the effect of age on CC shapes. The manifold considered is the same as that introduced in Section \ref{sec:sim}, but in this case $\sigma = 0.1$. Again the Levi-Civita connection is used.

\begin{figure}
\centering
\begin{minipage}{0.5\textwidth}
    \includegraphics[scale = 0.4, trim = 15 10 10 10, clip]{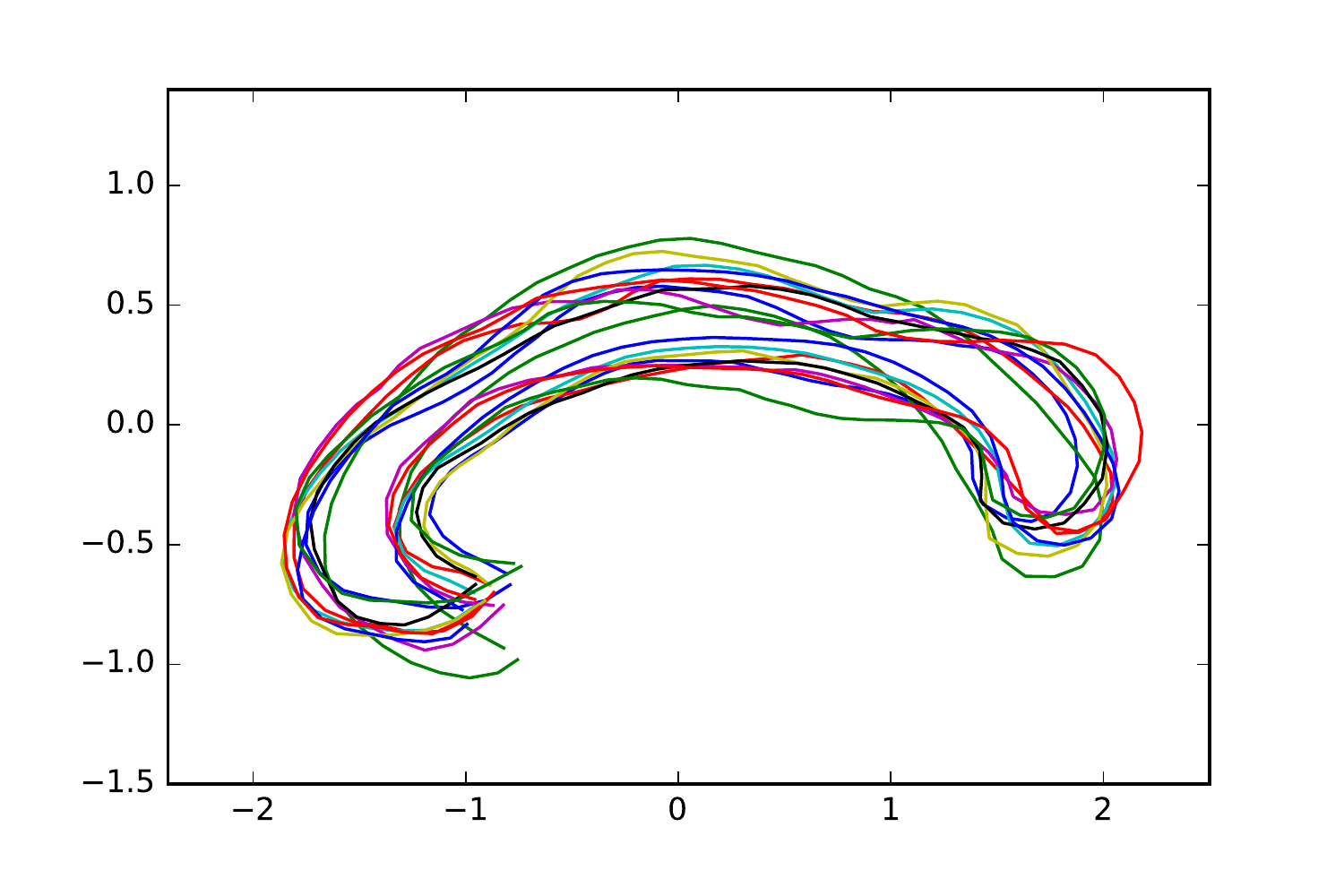}
\end{minipage}%
\begin{minipage}{0.5\textwidth}
    \includegraphics[scale = 0.4, trim = 15 10 10 10, clip]{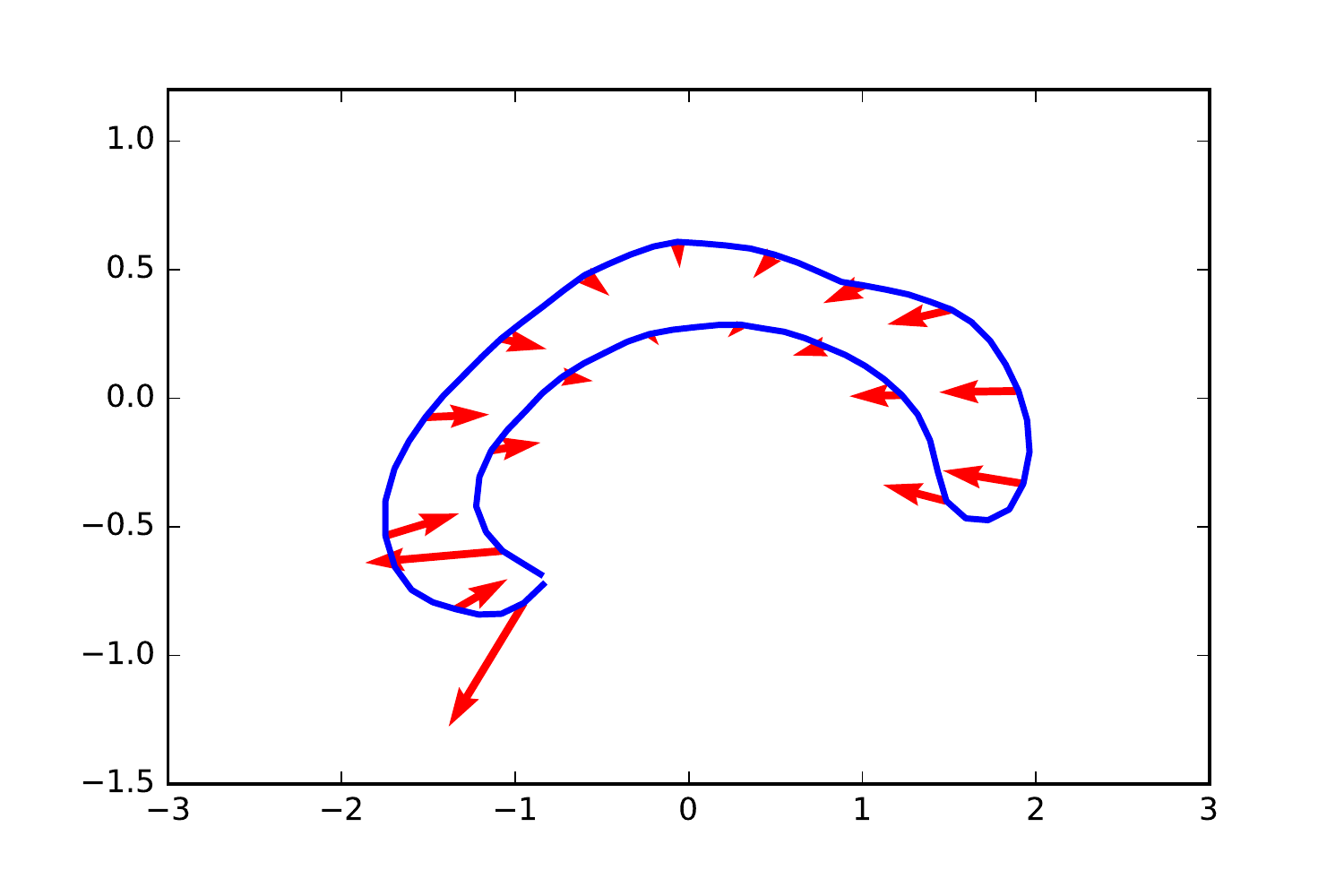}
\end{minipage}
\caption{(left) A subset of the Corpus Callosum data. (right) The mean shape with the estimated frame for the $20$ landmarks used in the model fitting.}
\label{fig:CCFrame}
\end{figure}

 A subset of the CC dataset is plotted in Figure \ref{fig:CCFrame}. For model fitting, a dataset of $20$ CC shapes was considered with age values ranging from $22$ to $78$. The model was fitted to CC shapes represented by a subset of $20$ landmarks. We did not incorporate a drift term in the model, and only the frame and $\tilde{W}$ has been estimated. The refrence point was set to the mean shape (Figure \ref{fig:CCFrame}) and $\tau = 0.1$.

 The estimated frame for the $20$ landmarks are shown in Figure \ref{fig:CCFrame} on top of the mean shape. The weight matrix was estimated as $\tilde{W} = -0.0002$. Given the low estimate of $\tilde{W}$ and hence a small frame matrix $W$, the result of this experiment suggests a low age effect on CC for these data.  

\section{Discussion}
\label{sec:Dis}

A method was proposed for modeling the relation between a manifold-valued response and Euclidean explanatory variables. The relation was modeled by transport of stochastic paths from $\R^m$ to the manifold. The stochastic paths defined on $\R^m$ was given as solutions to a stochastic differential equation with a contribution from a fixed drift, a stochastic process related to the explanatory variables, and a random noise assumed to follow a multidimensional Brownian motion. The response variable was then modeled as a noisy observation of a stochastic variable following the distribution of the endpoints of the transported process. The proposed model is intrinsic and based on a connection on the manifold without making linearization of the non-linear space. Moreover, a likelihood based estimation procedure were described using Laplace approximation of the marginal likelihood. We experimentally illustrated the model and the parameter estimation using a simulation study and a real data example.
 
Other procedures could be used for estimation of parameters. As an example, the Monte Carlo EM procedure could be used to optimize the complete data likelihood based on simulations of the missing data. Another example is to approximate the distribution of the response by moment matching.

An interesting problem to investigate is how to make variable selection in the model. As the contribution from the explanatory variables is defined in comparison with the frame basis vectors, one idea is to exclude those explanatory variables which corresponds to frame vectors parallel to the curve. These frame vectors will not contribute to the stochastic development and hence will not be important for explaining the relation to the response variable.

 An important assumption of the manifold considered, is that the manifold is equipped with a connection. In this paper, the Levi-Civita connection was used, but several other connections could have been chosen. It would be interesting to explore how the choice of connection affects the model.

As it is possible to transport stochastic paths from a manifold to a Euclidean space, the model could be generalized to handle situations in which a Euclidean response variable is compared to manifold-valued explanatory variables. Based on such a model, one might be able to make categorization of individuals based on manifold-valued shapes.



\bibliographystyle{plain}


\end{document}